\documentclass[amsmath,12pt]{article}

\begin{document}

\begin{titlepage}

\begin{center}

{\Large \bf
Classical relativistic ideal gas in thermodynamic equilibrium in a uniformly accelerated reference frame}
\vskip .6in

Domingo J. Louis-Martinez
\vskip .2in

Department of Physics and Astronomy,\\ University of British
Columbia\\Vancouver, Canada, V6T 1Z1 

martinez@phas.ubc.ca

\end{center}
 
\vskip 3cm
 
\begin{abstract}
A classical (non-quantum-mechanical) relativistic ideal gas in thermodynamic equilibrium in a uniformly 
accelerated frame of reference is studied using Gibbs's microcanonical and grand canonical formulations of statistical
mechanics.
Using these methods explicit expressions for the particle, energy and entropy density distributions are obtained, 
which are found to be in agreement with the well known results of the relativistic formulation of Boltzmann's kinetic theory. 
Explicit expressions for the total entropy, total energy and rest mass of the gas are obtained. The position of the center
of mass of the gas in equilibrium is found.
The non-relativistic and ultrarelativistic approximations are also considered. The phase space volume
of the system is calculated explicitly in the ultrarelativistic approximation. 
\end{abstract}


\end{titlepage}

{\bf Introduction:}

The relativistic ideal gas in a gravitational field has been extensively studied using the relativistic Boltzmann equation 
\cite{Tauber, Chernikov, Israel, Stewart, Ehlers, Groot, Cercignani}. 
This theory has found interesting applications in astrophysics, cosmology and nuclear physics \cite{Groot, Cercignani}.

General relativity is a local theory. The relativistic Boltzmann distribution function too is defined locally. It is therefore not
surprising that Boltzmann's kinetic theory has been the method employed in the study of relativistic gases in gravitational fields.

Gibbs's microcanonical formulation of statistical mechanics deals with systems of finite volumes and finite energies. 
To our knowledge, the methods of statistical mechanics, which are based on the microcanonical formulation, have not been employed
so far to check the validity of the results of the relativistic kinetic theory for a gas in the presence of a gravitational field. 
It seems of interest to check if both methods give the same results.

In section 1 we study a classical relativistic ideal gas in thermodynamic equilibrium in a uniformly accelerated 
reference frame  using Gibbs's microcanonical and grand canonical formulations 
of statistical mechanics.

In section 2 we compare the results of section 1 with the well known results of the relativistic kinetic theory. We find complete
agreement. 

In section 3 we show that the expressions for the particle, energy and entropy density distributions can be integrated
and we obtain explicit expressions for the total entropy, total energy and rest mass of the gas. We also find the center of mass of the system
in equilibrium and derive the condition for the relativistic gas to be in mechanical equilibrium in the uniformly accelerated frame.

The non-relativistic approximation is considered in the first part of section 4. The approximate expressions for the total energy and total entropy are found 
to be in agreement with known results for a classical non-relativistic ideal gas in a uniform gravitational field.

The ultrarelativistic approximation is studied in the second part of section 4. Explicit expressions are found in this case using the microcanonical
formalism. In particular, the phase space volume of the system can be found explicitly in this case, which allows a direct comparison
between the results of the microcanonical and the grand canonical formulations.

{\bf Section 1: The microcanonical and grand canonical formulations}

Let us consider a gas consisting of a very large number $N$ of identical (indistinguishable) structureless particles. Assume 
the particles of the gas do not interact with each other, except for elastic collisions among themselves and with the walls
of the container (an ideal gas). Let us assume the gas to be adiabatically isolated: any changes in the total internal energy
can only be the result of a change in the external parameters. We assume the container is in uniformly accelerated motion along
the $z$-direction with respect to an inertial reference frame $K'$, such that the components of the four-vector acceleration
$\ddot{x'}_{o}^{\mu}$ ($\mu = 0,1,2,3$) of the container obey the condition:

\begin{equation}
\ddot{x'}_{o}^{2} = \eta_{\mu\nu} \ddot{x'}_{o}^{\mu} \ddot{x'}_{o}^{\nu} = - \frac{g^{2}}{c^{2}} < 0,
\label{1}
\end{equation}

\noindent where $g$ is a constant, $\eta_{\mu\nu} = diag(+1,-1,-1,-1)$ and $c$ is the speed of light.

For simplicity, we assume the container to be a cylinder of base area A and height L. The axis of the cylinder is aligned along the 
$z$-direction.

In the local reference frame $K$ of the accelerated observer, with respect to whom the box is at rest, the metric tensor
can be written as \cite{Moller, MTW} \footnote{In the non-relativistic limit this metric takes the form:
$g_{\mu\nu} = diag\left(1+ \frac{2gz}{c^{2}}, -1, -1, -1\right)$, which is the expression for
the metric describing a uniform gravitational field in this approximation\cite{Moller}.}:

\begin{equation}
g_{\mu\nu} = diag\left(\left(1+ \frac{gz}{c^{2}}\right)^{2}, -1, -1, -1\right).
\label{2}
\end{equation} 

We assume the cylinder containing the gas is in a region within a distance $\frac{c^{2}}{g}$ of the observer. In other words, we
assume the container is small enough and close enough to the origin of the reference frame $K$ that the metric (\ref{2}) is valid
everywhere inside the container.

The Lagrangian for a system of $N$ non-interacting identical relativistic particles in the non-inertial uniformly accelerated
reference frame $K$ can be written as:

\begin{equation}
L = - m c^{2} \sum\limits_{a=1}^{N} \left(\left(1 + \frac{g z_{a}}{c^{2}}\right)^{2} 
- \frac{v_{a}^{2}}{c^{2}}\right)^{\frac{1}{2}},
\label{3}
\end{equation}

\noindent where $v_{a}^{i}$ ($i=1,2,3$) are the components of the velocity of the $a$th particle in $K$, 
$v_{a}^{2} = v_{a x}^{2} + v_{a y}^{2} + v_{a z}^{2}$, and $m$ is the mass of a particle.

From the Lagrangian (\ref{3}) we can find the components of the momentum of the $a$th particle:

\begin{equation}
p_{a i} = \frac{m v_{a}^{i}}{\left(\left(1 + \frac{g z_{a}}{c^{2}}\right)^{2} 
- \frac{v_{a}^{2}}{c^{2}}\right)^{\frac{1}{2}}},
\label{4}
\end{equation}

The Hamiltonian of the system can therefore be written as:

\begin{equation}
H = m c^{2} \sum\limits_{a=1}^{N} \left(1 + \frac{g z_{a}}{c^{2}}\right) 
\left(1 + \frac{p_{a}^{2}}{m^{2} c^{2}}\right)^{\frac{1}{2}},
\label{5}
\end{equation}

\noindent where $p_{a}^{2} = p_{a x}^{2} + p_{a y}^{2} + p_{a z}^{2}$.

Let us assume the gas has reached thermodynamic equilibrium. It is assumed that any infinitesimally small portion of the fluid
contains a very large number of particles.

In the non-inertial reference frame $K$, where all the portions of the fluid are at rest, the total energy of the fluid can
be written as:

\begin{equation}
E = A \int\limits_{0}^{L} \epsilon(z) dz,
\label{6}
\end{equation}

\noindent where $\epsilon(z)$ can be interpreted as the energy density of the fluid measured with respect to $K$.

In the instantaneous (freely falling) proper inertial frame $K'$, in which the fluid is also at rest, the total energy of the
fluid can be written as follows:

\begin{equation}
E' = A \int\limits_{0}^{L} \epsilon'(z') dz' = A \int\limits_{0}^{L} \epsilon'(z) dz,
\label{7}
\end{equation}

\noindent where $\epsilon'$ is the proper energy density (energy per unit proper volume) measured with respect to $K'$. 
Notice that both reference frames $K'$ and $K$ are at rest with respect to the fluid and with respect to each other, and
the metric tensor in $K$ is of the form (\ref{2}), therefore $dx' = dx$, $dy' = dy$ and $dz' = dz$. The dimensions of the container are
the same in both frames of reference. 

In $K'$ the particles of the gas move in rectilinear uniform motion in between collisions (as free particles), while in $K$
they are subject to the effect of non-inertial forces. The two quantities $\epsilon$ and $\epsilon'$ are related as follows:

\begin{equation}
\epsilon = \left(1 + \frac{gz}{c^{2}}\right) \epsilon'.
\label{8}
\end{equation}

The energy density $\epsilon$ of the gas (measured with respect to $K$) takes into account the effect of the non-inertial
forces on the particles, while the proper energy density $\epsilon'$ does not.

In the non-inertial reference frame $K$, where all portions of the fluid are at rest, the components of the energy-momentum tensor
of the fluid can be expressed as follows:

\begin{equation}
T^{\mu\nu} = diag\left(\frac{\epsilon'}{\left(1 + \frac{gz}{c^{2}}\right)^{2}}, p', p', p'\right),
\label{9}
\end{equation}

\noindent where $p'$ is the pressure in the fluid (a local function of position in the fluid) measured with respect to the instantaneous
inertial rest frame $K'$.

Equating the covariant derivative of the energy-momentum tensor to zero (law of energy-momentum conservation) we obtain the condition \cite{Weinberg, Landau5}
for the fluid to be in a state of mechanical equilibrium:

\begin{equation}
\frac{d p'}{dz} = - \frac{g}{c^{2}} \frac{(\epsilon' + p')}{\left(1 + \frac{gz}{c^{2}}\right)}.
\label{10}
\end{equation}

In the microcanonical formalism, the number $\Omega(E,N,g,A,L) = D(E,N,g,A,L) \delta E$ of accessible  microstates of the system  (for a 
fixed number of particles $N$ and fixed external parameters $g$, $A$ and $L$, and energy in the range $(E, E + \delta E)$) can be found
from the phase space volume integral:

\begin{eqnarray}
\Phi(E,N,g,A,L) & = &\int...\int dx_{1}dy_{1}dz_{1}dp_{1x}dp_{1y}dp_{1z}...dx_{N}dy_{N}dz_{N}dp_{Nx}dp_{Ny}dp_{Nz} \nonumber\\
& &  \sum\limits_{a=1}^{N} \left(1 + \frac{g z_{a}}{c^{2}}\right) 
\left(1 + \frac{p_{a}^{2}}{m^{2} c^{2}}\right)^{\frac{1}{2}} \le \frac{E}{m c^{2}}
\label{A}
\end{eqnarray}

\noindent as follows:

\begin{equation}
\Omega(E,N,g,A,L) = \frac{1}{N! h^{3N}}\left(\frac{\partial \Phi(E,N,g,A,L)}{\partial E}\right)_{N,g,A,L} \delta E.
\label{B}
\end{equation}

The phase space volume integral (\ref{A}) is the volume in phase space enclosed by the surface:

$ \sum\limits_{a=1}^{N} \left(1 + \frac{g z_{a}}{c^{2}}\right) 
\left(1 + \frac{1}{m^{2} c^{2}}\left(p_{a x}^{2} + p_{a y}^{2} + p_{a z}^{2}\right)\right)^{\frac{1}{2}} = \frac{E}{m c^{2}}$.

In (\ref{B}), $h$ is Planck's constant and the $N!$ accounts for the fact that the particles of the gas are indistinguishable.

From (\ref{A})  we immediately find:

\begin{eqnarray}
\Phi(E,N,g,A,L) & = & \left(4 \pi A\right)^{N} \int...\int p_{1}^{2}...p_{N}^{2} dz_{1}...dz_{N} dp_{1}...dp_{N} \nonumber\\
& &  \sum\limits_{a=1}^{N} \left(1 + \frac{g z_{a}}{c^{2}}\right) 
\left(1 + \frac{p_{a}^{2}}{m^{2} c^{2}}\right)^{\frac{1}{2}} \le \frac{E}{m c^{2}}. 
\label{C}
\end{eqnarray}

Assuming that the number of particles $N$ is very large ($N! \approx \sqrt{2 \pi N} \left(\frac{N}{e}\right)^{N}$ 
(Stirling's approximation)), the entropy of the isolated system in thermodynamic equilibrium
can be found from (\ref{B}) and (\ref{C}) as:

\begin{equation}
S = k \ln \Omega(E,N,g,A,L),
\label{D}
\end{equation}

\noindent where $k$ is Boltzmann's constant.

The quantities $T$ and $\mu$ can be found from the partial derivatives of the entropy as follows:

\begin{equation}
\frac{1}{T} = \left(\frac{\partial S}{\partial E}\right)_{N,g,A,L},
\label{E}
\end{equation}

\begin{equation}
\frac{\mu}{T} = - \left(\frac{\partial S}{\partial N}\right)_{E,g,A,L}.
\label{F}
\end{equation}

Let us consider a layer of the fluid (of area $A$) at a height $z$. This is an open subsystem. For this portion of the gas we can write the
grand canonical partition function:

\begin{eqnarray}
\mathcal{Z} & = &\sum\limits_{N=0}^{\infty} e^{\frac{\mu N}{kT}} \frac{(A dz)^{N}}{h^{3N} N!} 
\int...\int e^{-\sum\limits_{a=1}^{N} \frac{m c^{2}}{kT} \left(1 + \frac{gz}{c^{2}}\right) \left(1 + \frac{p_{a}^{2}}{m^{2} c^{2}}\right)^{\frac{1}{2}}}
d^{3}p_{1}...d^{3}p_{N} \nonumber\\
& = & \sum\limits_{N=0}^{\infty} e^{\frac{\mu N}{kT}} \frac{(A dz)^{N}}{h^{3N} N!} \left(4\pi 
\int\limits_{0}^{\infty} e^{- \frac{m c^{2}}{kT} \left(1 + \frac{gz}{c^{2}}\right) \left(1 + \frac{p^{2}}{m^{2}c^{2}}\right)^{\frac{1}{2}}} p^{2} dp\right)^{N}
\nonumber\\
& = & \sum\limits_{N=0}^{\infty} \frac{1}{N!} \left(e^{\frac{\mu}{kT}} 4\pi A dz \left(\frac{mc}{h}\right)^{3} 
\int\limits_{0}^{\infty} e^{- a(z) \left(1 + x^{2}\right)^{\frac{1}{2}}} x^{2} dx\right)^{N} \nonumber\\
& = & \exp\left(e^{\frac{\mu}{kT}} 4\pi A dz \left(\frac{mc}{h}\right)^{3} \int\limits_{0}^{\infty} e^{- a(z) \left(1 + x^{2}\right)^{\frac{1}{2}}} x^{2} dx\right),
\label{11}
\end{eqnarray}

\noindent where:

\begin{equation}
a(z) = \frac{m c^{2}}{kT} \left(1 + \frac{gz}{c^{2}}\right).
\label{12}
\end{equation} 

We notice that:

\begin{equation}
\int\limits_{0}^{\infty} e^{- a \left(1 + x^{2}\right)^{\frac{1}{2}}} x^{2} dx = \int\limits_{1}^{\infty} u \sqrt{u^{2} - 1} e^{- au} du = \frac{K_{2}(a)}{a},
\label{13}
\end{equation}

\noindent where $K_{2}$ is the modified Bessel function \cite{Abramowitz, Gradshteyn}.

From (\ref{11}), (\ref{12}) and (\ref{13}) we finally obtain the grand canonical partition function for the relativistic ideal gas:

\begin{equation}
\mathcal{Z} = \exp\left(e^{\frac{\mu}{kT}} 4\pi A dz \left(\frac{mc}{h}\right)^{3} 
\frac{K_{2}\left(\frac{m c^{2}}{kT} \left(1 + \frac{gz}{c^{2}}\right)\right)}{\frac{m c^{2}}{kT} 
\left(1 + \frac{gz}{c^{2}}\right)}\right).
\label{14}
\end{equation}

The grand canonical potential per unit volume in $K$, can be found directly from (\ref{14}):

\begin{equation}
\omega = - k e^{\frac{\mu}{kT}} 4\pi \left(\frac{mc}{h}\right)^{3} 
\frac{K_{2}\left(\frac{m c^{2}}{kT} \left(1 + \frac{gz}{c^{2}}\right)\right)}{\frac{m c^{2}}{kT} \left(1 + \frac{gz}{c^{2}}\right)}.
\label{15}
\end{equation}

From (\ref{15}) the entropy density $s$ (entropy per unit volume) and the particle density $n$ (number of particles per unit volume)
can be found using the equations:

\begin{equation}
s = - \left(\frac{\partial \omega}{\partial T}\right)_{g,\mu},
\label{16}
\end{equation}

\begin{equation}
n = - \left(\frac{\partial \omega}{\partial \mu}\right)_{g,T}.
\label{17}
\end{equation}

Using the thermodynamic identity:

\begin{equation}
\omega = \epsilon - Ts -\mu n,
\label{18}
\end{equation}

\noindent and expressing $\mu$ in terms of the particle density $n(0)$ at the bottom of the container, we obtain from (\ref{15}), (\ref{16}), (\ref{17}) and (\ref{18})
the particle density, entropy density, and energy density equilibrium distributions:

\begin{equation}
n(z) = n(0) \frac{K_{2}\left(\frac{m c^{2}}{kT} \left(1 + \frac{gz}{c^{2}}\right)\right)}{\left(1 + \frac{gz}{c^{2}}\right) K_{2}\left(\frac{m c^{2}}{kT}\right)},
\label{19}
\end{equation}

\begin{equation}
s(z) = n(z)k \left[\ln\left(\frac{4\pi e^{2}}{n(z)} \left(\frac{mc}{h}\right)^{3} 
\frac{K_{2}\left(\frac{m c^{2}}{kT} \left(1 + \frac{gz}{c^{2}}\right)\right)}{\frac{m c^{2}}{kT} \left(1 + \frac{gz}{c^{2}}\right)}\right) 
- \frac{m c^{2}}{kT} \left(1 + \frac{gz}{c^{2}}\right) 
\frac{K'_{2}\left(\frac{m c^{2}}{kT} \left(1 + \frac{gz}{c^{2}}\right)\right)}{K_{2}\left(\frac{m c^{2}}{kT} \left(1 + \frac{gz}{c^{2}}\right)\right)}\right], 
\label{20}
\end{equation}

\begin{equation}
\epsilon (z) = n(z) kT \left(1 - \frac{m c^{2}}{kT} \left(1 + \frac{gz}{c^{2}}\right) 
\frac{K'_{2}\left(\frac{m c^{2}}{kT} \left(1 + \frac{gz}{c^{2}}\right)\right)}{K_{2}\left(\frac{m c^{2}}{kT} \left(1 + \frac{gz}{c^{2}}\right)\right)}\right), 
\label{21}
\end{equation}

\noindent where $0\le z \le L$ (we assume the origin of the reference frame $K$ to be at the bottom of the container).

The pressure in $K$ can be found directly from the expression of the grand canonical potential per unit volume as follows:

\begin{equation}
p = - \omega = n(z) kT
\label{alpha}
\end{equation}

From (\ref{21}) and (\ref{8}) we find the proper energy density as a function of $z$:

\begin{equation}
\epsilon' (z) = \frac{n(z) kT}{\left(1 + \frac{gz}{c^{2}}\right)} 
\left(1 - \frac{m c^{2}}{kT} \left(1 + \frac{gz}{c^{2}}\right) 
\frac{K'_{2}\left(\frac{m c^{2}}{kT} \left(1 + \frac{gz}{c^{2}}\right)\right)}{K_{2}\left(\frac{m c^{2}}{kT} 
\left(1 + \frac{gz}{c^{2}}\right)\right)}\right).
\label{22}
\end{equation}

Substituting (\ref{22}) into (\ref{10}), we can solve the differential equation for the pressure $p'$ in $K'$ to obtain:

\begin{equation}
p'(z) = \frac{n(z) kT}{\left(1 + \frac{gz}{c^{2}}\right)}. 
\label{23}
\end{equation}

The thermodynamic relations between $\epsilon'$, $n'$, $s'$, $p'$ and the local temperature $T'$ measured in the 
instantaneous inertial frame $K'$ for a layer of gas in equilibrium at height $z$ can be found from
the grand canonical partition function (in the inertial frame):

\begin{eqnarray}
\mathcal{Z'} & = &\sum\limits_{N=0}^{\infty} e^{\frac{\mu'(z) N}{kT'(z)}} \frac{(A dz)^{N}}{h^{3N} N!} 
\int...\int e^{-\sum\limits_{a=1}^{N} \frac{m c^{2}}{kT'(z)} \left(1 + \frac{p_{a}^{\prime 2}}{m^{2} c^{2}}\right)^{\frac{1}{2}}}
d^{3}p'_{1}...d^{3}p'_{N} \nonumber\\
& = & \exp\left(e^{\frac{\mu'(z)}{kT'(z)}} 4\pi A dz \left(\frac{mc}{h}\right)^{3} 
\frac{K_{2}\left(\frac{m c^{2}}{kT'(z)}\right)}{\frac{m c^{2}}{kT'(z)}}\right).
\label{J}
\end{eqnarray}

As before, we can find $s'$ and $n'$ from the equations:

\begin{equation}
s' = - \left(\frac{\partial \omega'}{\partial T'}\right)_{\mu'},
\label{K}
\end{equation}

\begin{equation}
n' = - \left(\frac{\partial \omega'}{\partial \mu'}\right)_{T'}.
\label{L}
\end{equation}

\noindent where $\omega'$ can be found from (\ref{J}):

\begin{equation}
\omega' = - k e^{\frac{\mu'}{kT'}} 4\pi \left(\frac{mc}{h}\right)^{3} 
\frac{K_{2}\left(\frac{m c^{2}}{kT'}\right)}{\frac{m c^{2}}{kT'}}.
\label{M}
\end{equation}

Finally, using the thermodynamic identity:

\begin{equation}
\omega' = \epsilon' - T's' -\mu' n',
\label{N}
\end{equation}

\noindent and the relation:

\begin{equation}
\omega' = - p',
\label{O}
\end{equation}

\noindent for a relativistic ideal gas (in the inertial frame $K'$), in the absence of external forces, one finds the Juttner relations 
\cite{Juttner, Pauli, Synge, Anderson, Greiner, Hakim, Tecotl}:

\begin{equation}
\epsilon' = p' \left(1 - \frac{m c^{2}}{kT'} 
\frac{K'_{2}\left(\frac{m c^{2}}{kT'}\right)}{K_{2}\left(\frac{m c^{2}}{kT'}\right)}\right).
\label{24}
\end{equation}

\begin{equation}
p' = n' k T'
\label{25}
\end{equation}

\begin{equation}
s' = n' k \left[\ln\left(\frac{4\pi e^{2}}{n'} \left(\frac{mc}{h}\right)^{3} 
\frac{K_{2}\left(\frac{m c^{2}}{kT'}\right)}{\frac{m c^{2}}{kT'}}\right) 
- \frac{mc^{2}}{k T'}  
\frac{K'_{2}\left(\frac{m c^{2}}{kT'}\right)}{K_{2}\left(\frac{m c^{2}}{k T'}\right)}\right], 
\label{26}
\end{equation}

Comparing (\ref{24}), (\ref{25}) and (\ref{26}) with (\ref{22}), (\ref{23}) and (\ref{20}) we obtain:

\begin{equation}
T'(z) = \frac{T}{\left(1 + \frac{gz}{c^{2}}\right)},
\label{27}
\end{equation}

\begin{equation}
n'= n,
\label{28}
\end{equation}

\begin{equation}
s' =s.
\label{29}
\end{equation}

Equations (\ref{28}) and (\ref{29}) can also be obtained by noticing that the number of particles and the entropy are invariant quantities
(have the same values both in $K$ and $K'$) and the volumes in $K$ and $K'$ are the same (as noted above).

Equation (\ref{27}) is the well known Tolman relation \cite{Tolman, MTW} ($T' \sqrt{g_{00}} = T = const$) between the local temperature $T'(z)$ measured with
respect to the inertial frame $K'$ and the constant $T$ defined in the non-inertial frame $K$. We have arrived at (\ref{27}) using
the methods of statistical mechanics.

{\bf Section 2: The relativistic kinetic theory}

One can also obtain Eqs. (\ref{19} - \ref{21}) from the kinetic theory, using the relativistic Boltzmann equation 
\cite{Tauber, Chernikov, Israel, Stewart, Ehlers, Groot, Cercignani}.

In the non-inertial uniformly accelerated reference frame $K$, the relativistic Boltzmann equation can be written as follows:

\begin{eqnarray}
& &\frac{\partial f\left(t, \vec{r}, \vec{p}\right)}{\partial t} + 
\frac{\left(1 + \frac{g z}{c^{2}}\right)}{m \left(1 + \frac{p^{2}}{m^{2} c^{2}}\right)^{\frac{1}{2}}} 
\left(\vec{p} \frac{\partial f\left(t, \vec{r}, \vec{p}\right)}{\partial \vec{r}}\right)
- mg \left(1 + \frac{p^{2}}{m^{2} c^{2}}\right)^{\frac{1}{2}} \frac{\partial f\left(t, \vec{r}, \vec{p}\right)}{\partial p_{z}} \nonumber\\
& = & \frac{1}{h^{3}} \int \omega\left(\vec{r}; \vec{p}, \vec{p}_{1}; \vec{\tilde{p}}, \vec{\tilde{p}}_{1}\right) 
\left(f\left(t, \vec{r}, \vec{\tilde{p}}\right) f\left(t, \vec{r}, \vec{\tilde{p}}_{1}\right) - f\left(t, \vec{r}, \vec{p}\right) f\left(t, \vec{r}, \vec{p}_{1}\right)\right)
d^{3}p_{1}d^{3}\tilde{p}d^{3}\tilde{p}_{1} .\nonumber\\
\label{Z1}
\end{eqnarray}

The right hand side of the relativistic Boltzmann equation (\ref{Z1}) is the collision integral. It gives the rate of change, due to collisions,
in the mean number of particles in the phase space volume $dxdydzdp_{x}dp_{y}dp_{z}$. The left hand side of (\ref{Z1}) is 
$\frac{df}{dt} = \frac{\partial f}{\partial t} + \{f,H\}$, where $\{,\}$ is a Poisson bracket. The particular expression in the left hand side of (\ref{Z1})
is obtained by inserting the Hamiltonian (\ref{5}) in the Poisson bracket.

For elastic collisions: 

\begin{equation}
\vec{p} + \vec{p}_{1} = \vec{\tilde{p}} + \vec{\tilde{p}}_{1},
\label{Z2}
\end{equation}

\begin{equation}
\left(1 + \frac{p^{2}}{m^{2} c^{2}}\right)^{\frac{1}{2}} + \left(1 + \frac{p_{1}^{2}}{m^{2} c^{2}}\right)^{\frac{1}{2}} =
\left(1 + \frac{\tilde{p}^{2}}{m^{2} c^{2}}\right)^{\frac{1}{2}} + \left(1 + \frac{\tilde{p}_{1}^{2}}{m^{2} c^{2}}\right)^{\frac{1}{2}}.
\label{Z3}
\end{equation}

The mean number of particles per unit volume is given by the equation:

\begin{equation}
n(t, \vec{r}) = \frac{1}{h^{3}} \int f \left(t, \vec{r}, \vec{p}\right) d^{3}p. 
\label{Z4}
\end{equation}

In equilibrium:

\begin{equation}
\frac{\partial f_{eq}}{\partial t} = 0,
\label{Z5}
\end{equation}

\begin{equation}
f_{eq}\left(\vec{r}, \vec{\tilde{p}}\right) f_{eq}\left(\vec{r}, \vec{\tilde{p}}_{1}\right) = f_{eq}\left(\vec{r}, \vec{p}\right) f_{eq}\left(\vec{r}, \vec{p}_{1}\right).
\label{Z6}
\end{equation}

In Eq (\ref{Z6}) it is assumed that the initial and final momenta obey the conditions (\ref{Z2}, \ref{Z3}).

In our case, due to the symmetry of the problem considered, we also have the following conditions on $f_{eq}$:

\begin{equation}
\frac{\partial f_{eq}}{\partial x} = \frac{\partial f_{eq}}{\partial y} = 0.
\label{Z7}
\end{equation}

From (\ref{Z1}), (\ref{Z5} - \ref{Z7}) we obtain that the distribution function $f_{eq}$ satisfies the equation:

\begin{equation}
\frac{\left(1 + \frac{g z}{c^{2}}\right) p_{z}}{m \left(1 + \frac{p^{2}}{m^{2} c^{2}}\right)^{\frac{1}{2}}} 
\frac{\partial f_{eq}\left(z, \vec{p}\right)}{\partial z}
- mg \left(1 + \frac{p^{2}}{m^{2} c^{2}}\right)^{\frac{1}{2}} \frac{\partial f_{eq}\left(z, \vec{p}\right)}{\partial p_{z}} = 0.
\label{Z8}
\end{equation}

The solution of (\ref{Z8}), which also satisfies the condition (\ref{Z6}) (assuming (\ref{Z2}, \ref{Z3})), can be written as:

\begin{equation}
f_{eq}\left(z, \vec{p}\right) = \alpha e^{- \beta mc^{2} \left(1 + \frac{gz}{c^{2}}\right) \left(1 + \frac{p^{2}}{m^{2} c^{2}}\right)^{\frac{1}{2}}},
\label{Z9}
\end{equation} 

\noindent where $\alpha$ and $\beta$ are constants.

Substituting (\ref{Z9}) into (\ref{Z4}) and performing the integration we obtain:

\begin{equation}
n(z) = \frac{4 \pi \alpha m^{2} c}{\beta} \frac{K_{2}\left(\beta m c^{2} \left(1 + \frac{gz}{c^{2}}\right)\right)}{\left(1 + \frac{gz}{c^{2}}\right)}.
\label{Z10}
\end{equation}

In terms of the particle density at the bottom of the container we obtain:

\begin{equation}
n(z) = \frac{n(0)}{K_{2}\left(\beta m c^{2}\right)} \frac{K_{2}\left(\beta m c^{2} \left(1 + \frac{gz}{c^{2}}\right)\right)}{\left(1 + \frac{gz}{c^{2}}\right)}.
\label{Z11}
\end{equation}

The energy density $\epsilon(z)$ in $K$ can be derived from the relation:

\begin{equation}
\epsilon(z) = \frac{1}{h^{3}} \int m c^{2} \left(1 + \frac{gz}{c^{2}}\right) \left(1 + \frac{p^{2}}{m^{2} c^{2}}\right)^{\frac{1}{2}} f_{eq} \left(z, \vec{p}\right) d^{3}p
\label{Z12}
\end{equation}

Substituting (\ref{Z9}) into (\ref{Z12}), taking into account (\ref{Z10}), and performing the integration we obtain:

\begin{equation}
\epsilon (z) = \frac{n(z)}{\beta} \left(1 - \beta m c^{2} \left(1 + \frac{gz}{c^{2}}\right) 
\frac{K'_{2}\left(\beta m c^{2} \left(1 + \frac{gz}{c^{2}}\right)\right)}{K_{2}\left(\beta m c^{2} \left(1 + \frac{gz}{c^{2}}\right)\right)}\right)
\label{Z13}
\end{equation}

Finally, we can obtain the entropy density from the relation \cite{Lifshitz}:

\begin{equation}
s(z) = \frac{k}{h^{3}} \int f_{eq} \ln\left(\frac{e}{f_{eq}}\right) d^{3}p.
\label{Z14}
\end{equation}

Substituting (\ref{Z9}) into (\ref{Z14}), taking into account (\ref{Z10}), and performing the integration we obtain:

\begin{equation}
s(z) = n(z)k \left[\ln\left(\frac{4\pi e^{2}}{n(z)} \left(\frac{mc}{h}\right)^{3} 
\frac{K_{2}\left(\beta m c^{2} \left(1 + \frac{gz}{c^{2}}\right)\right)}{\beta m c^{2} \left(1 + \frac{gz}{c^{2}}\right)}\right) 
- \beta m c^{2} \left(1 + \frac{gz}{c^{2}}\right) 
\frac{K'_{2}\left(\beta m c^{2} \left(1 + \frac{gz}{c^{2}}\right)\right)}{K_{2}\left(\beta m c^{2} \left(1 + \frac{gz}{c^{2}}\right)\right)}\right], 
\label{Z15}
\end{equation}

Comparing (\ref{Z11}, \ref{Z12}, \ref{Z15}) with (\ref{19} - \ref{21}) we find the constant $\beta$:

\begin{equation}
\beta = \frac{1}{kT}.
\label{Z16}
\end{equation}

{\bf Section 3: Global quantities}

The density of particles at the bottom of the container $n(0)$ can be related to the total number of particles in the system $N$ by direct 
integration of (\ref{19}):

\begin{equation}
N = A \int\limits_{0}^{L} n(z) dz = \frac{A n(0)}{K_{2}\left(\frac{m c^{2}}{kT}\right)} \frac{c^{2}}{g} \int\limits_{a(0)}^{a(L)} \frac{K_{2}(u)}{u} du
=  \frac{A n(0)}{K_{2}\left(\frac{m c^{2}}{kT}\right)} \frac{kT}{mg} \left(K_{1}\left(\frac{m c^{2}}{kT}\right) - 
\frac{K_{1}\left(\frac{m c^{2}}{kT}\left(1 + \frac{gL}{c^{2}}\right)\right)}{\left(1 + \frac{gL}{c^{2}}\right)}\right),
\label{30}
\end{equation}

\noindent where we used the relation: $\frac{K_{2}(u)}{u}= - \frac{d}{du} \left(\frac{K_{1}(u)}{u}\right)$ \cite{Abramowitz, Gradshteyn}.

The total energy of the system $E$ in $K$ can be found by substituting (\ref{19}) and (\ref{30}) into (\ref{21}) and performing the integration (\ref{6}):

\begin{equation}
E = NkT + Nm c^{2} \frac{\left(K_{2}\left(\frac{m c^{2}}{kT}\right) - 
K_{2}\left(\frac{m c^{2}}{kT}\left(1 + \frac{gL}{c^{2}}\right)\right)\right)}
{\left(K_{1}\left(\frac{m c^{2}}{kT}\right) - 
\frac{K_{1}\left(\frac{m c^{2}}{kT}\left(1 + \frac{gL}{c^{2}}\right)\right)}{\left(1 + \frac{gL}{c^{2}}\right)}\right)}.
\label{31}
\end{equation}

Similarly, substituting (\ref{19}) and (\ref{30}) into (\ref{20}), and integrating, we obtain the total entropy of the gas:

\begin{eqnarray}
S & = & Nk \ln\left(\frac{A}{N} 4\pi e^{2} \left(\frac{mc}{h}\right)^{3} \left(\frac{kT}{m c^{2}}\right)^{2} \frac{c^{2}}{g}
\left(K_{1}\left(\frac{m c^{2}}{kT}\right) - 
\frac{K_{1}\left(\frac{m c^{2}}{kT}\left(1 + \frac{gL}{c^{2}}\right)\right)}{\left(1 + \frac{gL}{c^{2}}\right)}\right)\right)
\nonumber\\
& & + \frac{Nm c^{2}}{T} \frac{\left(K_{2}\left(\frac{m c^{2}}{kT}\right) - 
K_{2}\left(\frac{m c^{2}}{kT}\left(1 + \frac{gL}{c^{2}}\right)\right)\right)}
{\left(K_{1}\left(\frac{m c^{2}}{kT}\right) - 
\frac{K_{1}\left(\frac{m c^{2}}{kT}\left(1 + \frac{gL}{c^{2}}\right)\right)}{\left(1 + \frac{gL}{c^{2}}\right)}\right)}.
\label{32}
\end{eqnarray}

The total energy of the system $E'$ in $K'$ can be found by substituting (\ref{19}) and (\ref{30}) into (\ref{22}) and performing the integration (\ref{7}):

\begin{equation}
E' = Nm c^{2} \frac{\left(K_{2}\left(\frac{m c^{2}}{kT}\right) - 
\frac{K_{2}\left(\frac{m c^{2}}{kT}\left(1 + \frac{gL}{c^{2}}\right)\right)}{\left(1 + \frac{gL}{c^{2}}\right)}\right)}
{\left(K_{1}\left(\frac{m c^{2}}{kT}\right) - 
\frac{K_{1}\left(\frac{m c^{2}}{kT}\left(1 + \frac{gL}{c^{2}}\right)\right)}{\left(1 + \frac{gL}{c^{2}}\right)}\right)}.
\label{39}
\end{equation}

$E'$ given by Eq(\ref{39}) is the proper energy of the gas.

The position of the center of mass of the gas with respect to the inertial frame $K'$ can be determined by the equation \cite{Landau2}:

\begin{equation}
z_{c.m.} = \frac{A \int\limits_{0}^{L} z \epsilon'(z) dz}{E'}
\label{40}
\end{equation}

The integration in (\ref{40}) can be easily carried out and using (\ref{39}) and (\ref{31}) one finds that:

\begin{equation}
Mg z_{c.m.} = E - E'.
\label{41}
\end{equation}

\noindent where the rest mass $M$ of the fluid is defined as:

\begin{equation}
M = \frac{E'}{c^{2}}
\label{42}
\end{equation}

From (\ref{42}), (\ref{39}), (\ref{alpha}), (\ref{19}) and (\ref{30}), we notice that:

\begin{equation}
Mg = A \left(p(0) - p(L)\right).
\label{43}
\end{equation}

The above equation can be interpreted as the condition (in K) for the object to be in static mechanical equilibrium: the net force exerted by the walls of the container 
on the gas as a whole equals its weight. Notice that in the non-inertial reference frame $K$ the net force exerted on the gas by the side walls of the cylinder 
is equal to zero.

{\bf Section 4: The non-relativistic and ultrarelativistic approximations}

{\it The non-relativistic approximation.}

In the non-relativistic limit ($\frac{m c^{2}}{kT} \to \infty$) we can use the expressions for the asymptotics of the modified Bessel 
functions \cite{Abramowitz}:

\begin{equation}
K_{\nu}(z) \approx \sqrt{\frac{\pi}{2 z}} e^{-z} \left(1 + \frac{4 \nu^{2} - 1}{8 z}\right),
\label{33}
\end{equation}

\noindent for $\nu = 1, 2$ in (\ref{31}) and (\ref{32}) to obtain the non-relativistic expressions for the total energy and total entropy of the gas:

\begin{equation}
E = Nm c^{2} + \frac{5}{2}NkT - \frac{NmgL}{\left(e^{\frac{mgL}{kT}} - 1\right)},
\label{34}
\end{equation} 

\begin{equation}
S = Nk \left(1 - \frac{mgL}{kT} \frac{1}{\left(e^{\frac{mgL}{kT}} - 1\right)}\right) +
Nk \ln\left(\frac{A}{N} \frac{(2\pi mkT)^{\frac{3}{2}}}{h^{3}} e^{\frac{5}{2}} \frac{kT}{mg} \left(1 - e^{- \frac{mgL}{kT}}\right)\right).
\label{35}
\end{equation}

Equations (\ref{34}) and (\ref{35}) are in perfect agreement with the results for the non-relativistic classical ideal gas in a uniform gravitational
field obtained in \cite{Landsberg}. 

In the non-relativistic limit, where $\left(1 + \frac{p_{a}^{2}}{m^{2} c^{2}}\right)^{\frac{1}{2}} \approx 1 + \frac{p_{a}^{2}}{2 m^{2}c^{2}}$, 
the integration in (\ref{C}) can also be carried out and (assuming $E > Nmc^{2} +NmgL$) we obtain results in 
agreement with \cite{Roman}.

{\it The ultrarelativistic approximation.}

In the ultrarelativistic approximation, where $\left(1 + \frac{p_{a}^{2}}{m^{2} c^{2}}\right)^{\frac{1}{2}} \approx \frac{p_{a}}{mc}$, 
the integration in (\ref{C}) can be carried out explicitly and (assuming $E > Nm c^{2}\left(1 + \frac{gL}{c^{2}}\right)$) we obtain:

\begin{equation}
\Phi(E,N,g,A,L) = \frac{1}{(3N)!} 
\left(\frac{4\pi A}{gc} E^{3} \left(1 - \frac{1}{\left(1 + \frac{gL}{c^{2}}\right)^{2}}\right)\right)^{N}.
\label{G}
\end{equation}

From (\ref{G}), (\ref{B}) and (\ref{D}), using Stirling's approximation, we obtain a formula for the entropy of an ultrarelativistic ideal gas in
thermodynamic equilibrium in a uniformly accelerated reference frame:

\begin{equation}
S = Nk \ln\left(\frac{4\pi e^{4}}{gc h^{3}} \frac{A}{N} \left(\frac{E}{3N}\right)^{3} 
\left(1 - \frac{1}{\left(1 + \frac{gL}{c^{2}}\right)^{2}}\right)\right).
\label{H}
\end{equation}

In the ultrarelativistic limit ($\frac{m c^{2}}{kT} \to 0$) we can use the approximate expressions \cite{Abramowitz, Gradshteyn} for the modified Bessel
functions as $z \to 0$:

\begin{equation}
K_{\nu}(z) \approx \frac{\Gamma(\nu)}{2} \left(\frac{2}{z}\right)^{\nu}
\label{36}
\end{equation}

\noindent for $\nu =1,2$, in (\ref{31}) and (\ref{32})  to obtain:

\begin{equation}
E = 3NkT
\label{37}
\end{equation}  

\begin{equation}
S = Nk \ln\left(\frac{4\pi e^{4}}{gc h^{3}} \frac{A}{N} \left(kT\right)^{3} 
\left(1 - \frac{1}{\left(1 + \frac{gL}{c^{2}}\right)^{2}}\right)\right).
\label{38}
\end{equation}  

Combining Eqs. (\ref{38}) and (\ref{37}) we obtain (\ref{H}), which shows that the microcanonical and the grand canonical formalisms
give identical results in the ultrarelativistic approximation.

In this paper we have considered the case of an ideal gas in a uniformly accelerated frame. The relativistic kinetic theory is applicable in the more general 
case of a gas in a stationary gravitational field. It would be of interest to generalize the results obtained here to find the microcanonical and grand canonical
formulations for the gas in a stationary gravitational field.

\end{document}